\documentclass[aps, prl, preprint, 10pt, twocolumn]{revtex4-1}
\usepackage{graphicx}
\usepackage{amssymb,amsfonts,amsmath}

\begin{document}
\title{Structure of a model salt bridge in solution investigated with
2D-IR spectroscopy}
\author{Adriana Huerta-Viga}
\author{S\'ergio R. Domingos}
\author{Saeed Amirjalayer}
\author{Sander Woutersen}

\email{S.Woutersen@uva.nl}

\affiliation{Van 't Hoff Institute for Molecular
Sciences (HIMS), University of Amsterdam\\
Science Park 904 Amsterdam, The Netherlands, 1098 XH}

\begin{abstract}
\noindent
Salt bridges are known to be important for the stability
of protein conformation, but up to now it has been difficult to study their
geometry in solution. Here we characterize the spatial structure of a model
salt bridge between guanidinium (Gdm$^+$) and acetate (Ac$^-$) using
two-dimensional vibrational (2D-IR) spectroscopy. We find that as a result of
salt bridging the infrared response of Gdm$^+$ and Ac$^-$ change significantly,
and in the 2D-IR spectrum, salt bridging of the molecules appears as cross
peaks. From the 2D-IR spectrum we determine the relative orientation of the
transition-dipole moments of the vibrational modes involved in the salt bridge,
as well as the coupling between them. In this manner we reconstruct the geometry
of the solvated salt bridge. 
\end{abstract}

\maketitle

\begin{figure}[!b]
  \begin{center}
  \includegraphics[width=8.5cm]{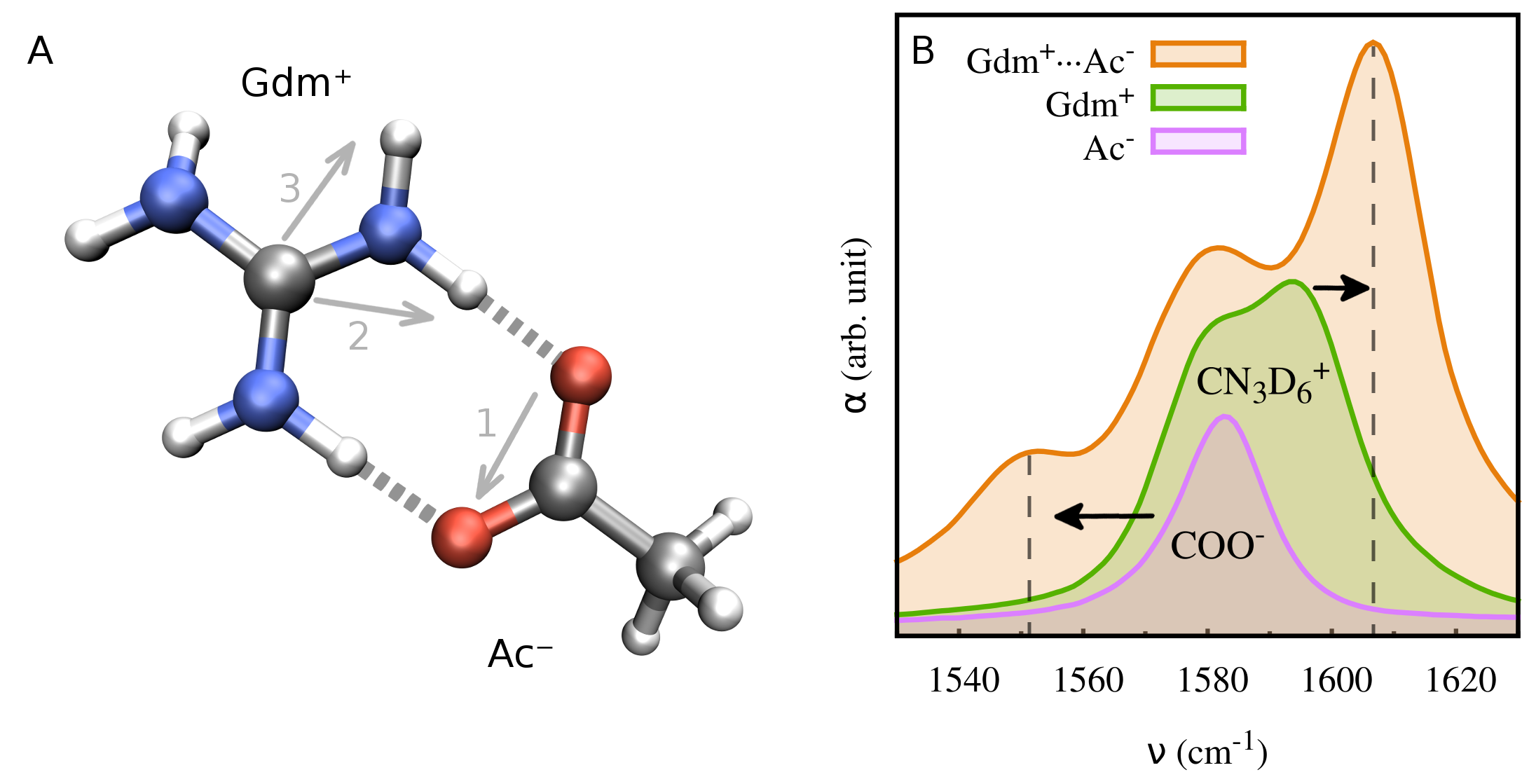}
  \end{center}
  \caption{(A) Molecular structure of the Gdm$^+\cdots$Ac$^-$ dimer obtained
using {\it ab initio} methods. The corresponding transition-dipole
moments of the COO$^-$ stretch mode (1) and of the
CN$_3$D$_6^+$ low and high frequency modes (2 and 3, respectively)  are
indicated by arrows. (B) Infrared absorption spectrum of Gdm$^+\!\cdots$Ac$^-$,
Gdm$^+$ and Ac$^-$ in DMSO (400 mM, solvent subtracted). Shifts of the COO$^-$
and high-frequency CN$_3$D$^+_6$ bands are indicated by arrows.}
 \label{fig:molecule-FTIR}
\end{figure}

The stabilization of molecular conformations by the attractive interaction
between oppositely charged ions (salt bridges) is of great relevance in
many areas of science. In particular, biological systems such as proteins,
often contain salt bridges between ionic side chains that determine their
structure~\cite{kumar1999} and function.~\cite{christie2012} 
It is therefore fundamental to characterize the solvated
structure of salt-bridged ion pairs, but this is unfortunately not possible with
conventional methods like NMR.  
In this communication we report the study of a biologically-relevant
ion pair, that formed by guanidinium (Gdm$^+$) and acetate
(Ac$^-$), using two-dimensional infrared (2D-IR) spectroscopy. This ion pair is
a model for salt bridges between an arginine and a carboxylate side group (from
glutamate
or aspartate), which occur commonly in
proteins.~\cite{walker2009} The molecular structure of this ion pair is shown in
Fig.\ \ref{fig:molecule-FTIR}A.
Isolated Gdm$^+$ has D$_3$ symmetry and a degenerate mode at 1600~cm$^{-1}$
due to a combined CN$_3$ antisymmetric stretch and NH$_2$ scissors
motion.~\cite{magalhaes1997} This degeneracy is also observed in aqueous
solution, but it is broken in viscous solvents.~\cite{vorobyev2010}
When dissolving deuterated Gdm$^+$ (guanidine$\cdot$DCl, $>$98\% purity) in
deuterated  dimethylsulfoxide (DMSO), we observe a similar splitting
between the frequencies of the two CN$_3$D$_6^+$ modes, as can be seen in
Fig.\ \ref{fig:molecule-FTIR}B. In the following, we will refer to the high-
and low-frequency CN$_3$D$_6^+$ of Gdm$^+$ as the Gdm$^+_{\rm HF}$ and
Gdm$^+_{\rm LF}$ modes, respectively.

\noindent
Interestingly, when an equimolar amount of
Ac$^-$ ions is added to the solution (guanidine acetate salt, $>$98\%
purity), this splitting becomes larger. It is known that Gdm$^+$ and Ac$^-$
have a
strong
binding affinity in DMSO, forming more than 98\% dimers at the concentration
used in our experiments.~\cite{linton1999} This suggests that the larger
splitting between the
Gdm$^+$ modes is due
to an interaction with the Ac$^-$ ion. Moreover, Ac$^-$ (tetrabutylammonium
acetate, $>$97\% purity) has an absorption band
at 1580~cm$^{-1}$ in DMSO (shown in Fig.\ \ref{fig:molecule-FTIR}B) due to the
COO$^-$ antisymmetric stretch mode. This mode red-shifts after
dimerization with
Gdm$^+$. 
The change in the infrared response of both the Gdm$^+$ and the Ac$^-$ ions
upon dimerization strongly suggests that there is a coupling between the
vibrational modes of these two molecules. 

\begin{figure}[ht]
  \begin{center}
  \includegraphics[width=8.5cm]{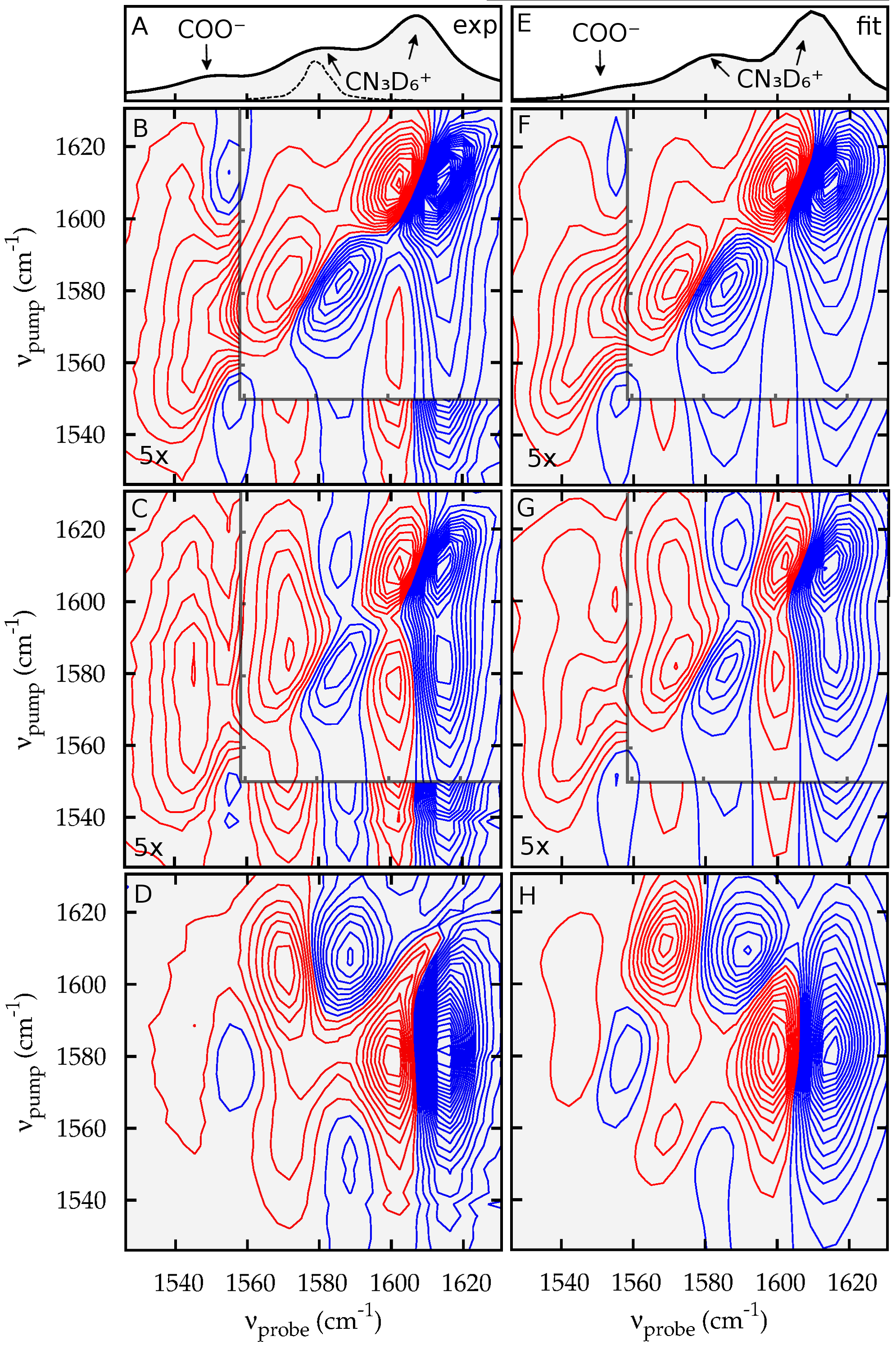}
  \end{center}
  \caption{(A) Infrared absorption spectrum of Gdm$^+\!\cdots$Ac$^-$.
The dashed line indicates a representative pump-pulse spectrum. (B) and (C)
2D-IR spectra of Gdm$^+\cdots$Ac$^-$ for parallel and perpendicular polarization
of the pump and probe pulses, respectively (pump-probe delay of 2 ps). Blue
indicates negative absorption change, red positive absorption change. The
contour intervals are 1 mOD (parallel) and 0.5 mOD (perpendicular), and the
left and bottom parts are scaled by a factor of five, as indicated. 
(D) Difference 2D-IR spectrum between parallel and three times perpendicular
polarization of the pump and probe pulses, the contour intervals are 0.5 mOD.
(E) Calculated absorption spectrum with parameters obtained from a fit
to the 2D-IR spectrum. (F) and (G) Fitted 2D-IR spectra for parallel and
perpendicular polarization of the pump and probe pulses, respectively. The
contour intervals are the same as in (B) and (C). (H) Calculated difference
2D-IR spectrum, with the same contour intervals as in (D).}
 \label{fig:2D}
\end{figure}

\begin{figure}[!b]
  \begin{center}
  \includegraphics[width=8.5cm]{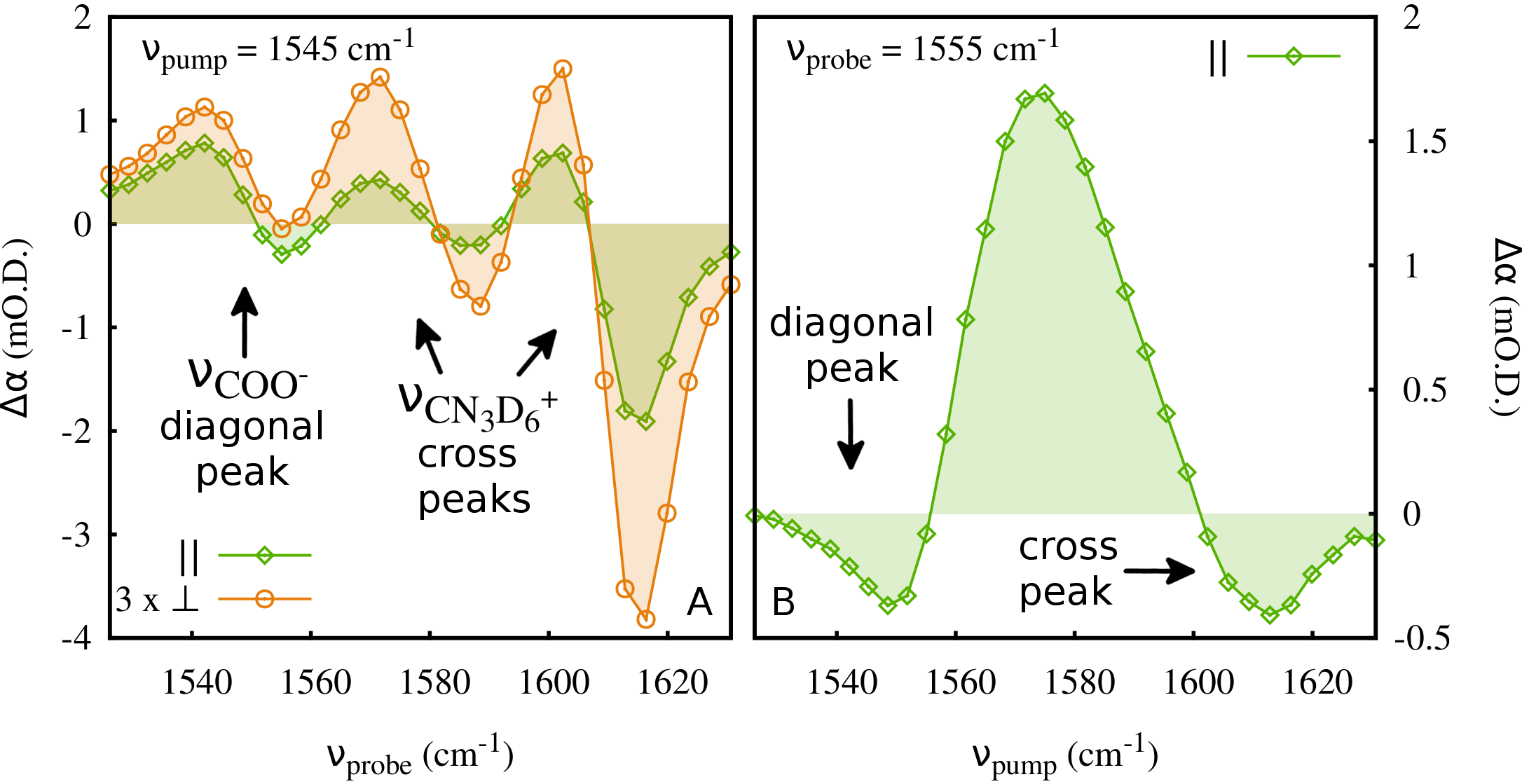}
  \end{center}
  \caption{(A) Cross section along the probe axis of the 2D-IR spectra for
parallel and perpendicular polarization of the pump and probe pulses, and for
$\nu_{\rm pump}=\nu_{\rm COO^-}$. (B) Cross
section along the pump axis of the 2D-IR spectrum for parallel polarization of
the pump and probe pulses, and for $\nu_{\rm probe}=\nu_{\rm COO^-}$.}
\label{fig:slices}
\end{figure}

\noindent
The 2D-IR  spectrum of Gdm$^+\cdots$Ac$^-$ confirms unambiguously that the
Gdm$^+_{\rm HF}$ and Gdm$^+_{\rm LF}$ modes are both coupled to the
COO$^-$ stretch mode of Ac$^-$. We use a femtosecond pump-probe setup that has
been described elsewhere.~\cite{huerta-viga10} The resulting
spectra are shown in Figs.\ \ref{fig:2D}B and C for parallel and perpendicular
polarization of the
pump and probe pulses, respectively. The non-zero off-diagonal response in the
2D-IR spectrum
indicates that there is a coupling between the two CN$_3$D$_6^+$ modes of
Gdm$^+$ and, more importantly, between each of them and the COO$^-$ stretch
mode of Ac$^-$. 
These cross peaks can be seen better in slices along both the
pump and probe axes of the 2D-IR spectra. Fig.\ \ref{fig:slices}A shows a
cross section for $\nu_{\rm pump}=\nu_{\rm COO^-}$ for parallel and
perpendicular polarization of the pump and probe pulses. In the cross section,
the cross peaks between the COO$^-$ stretch mode and each of the two
CN$_3$D$_6^+$
modes are clearly visible. Note that the diagonal response of the COO$^-$
stretch mode has a
smaller magnitude than these cross peaks because of its smaller absorption cross
section, see Fig.\ \ref{fig:molecule-FTIR}. 
Fig.\ \ref{fig:slices}B shows a cross section along
the probe axis for $\nu_{\rm probe}=\nu_{\rm COO^-}$ and for parallel
polarization of pump and probe pulses. The negative part at 1550~cm$^{-1}$ is
due to the
bleaching and $\nu=1\rightarrow 0$ stimulated emission of the COO$^-$ stretch
mode on the diagonal. The positive region centered
at 1580~cm$^{-1}$ is the low-probe-frequency tail of the diagonal induced
absorption of the Gdm$^+_{\rm LF}$ mode. The negative region at 1610~cm$^{-1}$
is the
negative part of the cross peak between the COO$^-$ stretch mode and the
Gdm$^+_{\rm HF}$ mode. 

In order to obtain structural information from the 2D-IR spectra, we model them
with an exciton model, which is a well established way of analyzing
2D-IR data.~\cite{hammzanni}  It has been shown before that, in addition to
coupling between the two CN$_3$D$_6^+$ modes, there
is also energy transfer between them,\cite{vorobyev2010} but in DMSO this
process is slow enough to be neglected at the delay at which we measured the
2D-IR spectra used for the structural analysis (2 ps).  
Figs.\ \ref{fig:2D}F and G show the calculated 2D-IR
spectra, which are in very good agreement with the measured ones, specially
considering that the difference 2D-IR spectrum between parallel and
perpendicular polarizations is reproduced very well (even though it is not
included in the fit). The couplings between the three modes involved in the salt
bridge
(Gdm$^+_{\rm HF}$, Gdm$^+_{\rm LF}$ and Ac$^-$), and the angles between their
transition-dipole
moments were parameters of the fit and are listed in Table
\ref{tbl:parameters}.  
The coupling between the Gdm$^+_{\rm HF}$ and Gdm$^+_{\rm LF}$ modes is
9~cm$^{-1}$, but a large
coupling was expected between these vibrational modes because they share atoms
and bonds. Remarkably, the coupling between each of the Gdm$^+_{\rm HF}$ and
Gdm$^+_{\rm LF}$ modes and the COO$^-$ stretch mode is also large, (10
and 7~cm$^{-1}$, respectively) even though these vibrational modes neither share
atoms nor bonds. This coupling rather originates from salt bridging between the
two molecules, an interaction that has both an electrostatic and hydrogen-bond
nature. The large couplings between the COO$^-$ and the Gdm$^+_{\rm HF, LF}$
modes
show that the large frequency shifts observed in the IR spectrum upon
salt-bridge formation (Fig.\ \ref{fig:molecule-FTIR}A) are mostly due to
splitting of the coupled modes rather than to a change in the local-mode
frequencies.  
The angles between the transition-dipole moments of the salt-bridged
vibrational modes are listed in Table \ref{tbl:parameters}, and these values
are  in agreement with a planar geometry.  The planarity of a salt bridge is
often taken as a metric of its quality in X-ray studies,~\cite{donald2011} and
our results seem to suggest that in DMSO solution, the geometry of an isolated
salt bridge, in which steric constraints are absent, is indeed planar. We
have performed complementary {\it ab initio} calculations on the
Gdm$^+\cdots$Ac$^-$ dimer using Gaussian03~\cite{g03} at MP2/6-311+G(d) level of
theory. The calculation predicts an approximately planar geometry for the salt
bridge, in which all three transition-dipole moments lie
almost in the same plane, as shown in Fig. \ref{fig:molecule-FTIR}A. The angle
between the transition-dipole vectors of the two Gdm$^+$ modes is smaller than
for isolated
guanidinium~\cite{vorobyev2010}, most likely as a result of salt bridging with
the Ac$^-$ ion. We find in the calculation that the Gdm$^+_{\rm HF}$ mode is
antisymmetric with respect to the symmetry axis through the C-C bond of Ac$^-$,
which explains the large coupling with the also antisymmetric COO$^-$
stretch mode.~\cite{sharma1981} The Gdm$^+_{\rm LF}$ mode is
symmetric with respect to this symmetry axis, so it is remarkable that
it also couples strongly to the COO$^-$ stretch mode, despite their different
symmetry.

\noindent
In conclusion, we were able to detect the existence of a salt bridge between
Gdm$^+$ and Ac$^-$ in solution using 2D-IR spectroscopy.  We characterize the
coupling between two CN$_3$D$_6^+$ modes of Gdm$^+$ and the COO$^-$ stretch
mode of Ac$^-$. We find that the COO$^-$ stretch mode couples more strongly to
the high-frequency CN$_3$D$_6^+$ mode than to the low-frequency one, most
likely because of their similar symmetry. We determine the geometry of the salt
bridge and find that it is in good agreement with a salt-bridge geometry in
which the Gdm$^+$ and COO$^-$ moieties are coplanar.

\begin{table}[!t]
\small
  \caption{\ Coupling $\beta$ and angle $\theta$ between the transition-dipole
moments of the COO$^-$ stretch of Ac$^-$ and the two CN$_3$D$_6^+$ modes of
Gdm$^+$ as obtained from the fit ($\chi^2_r=43$).}
  \label{tbl:parameters}
  \begin{center}
  \begin{tabular*}{0.47\textwidth}{@{\extracolsep{\fill}}lll}
    \hline
    Modes & $\beta$ (cm$^{-1}$) & $\theta$ \\
    \hline
    Gdm$_{\rm HF}^+$-Gdm$_{\rm LF}^+$ & 9 & 80$^\circ$\\
    Gdm$_{\rm HF}^+$-Ac$^-$ & 10 & -3$^\circ$\\
    Gdm$_{\rm LF}^+$-Ac$^-$ & 7 & 80$^\circ$\\
    \hline  
  \end{tabular*}
  \end{center}
\end{table}

\bibliography{huerta-viga2013} 
\bibliographystyle{unsrtnat}

\end{document}